\documentclass[journal,comsoc]{IEEEtran}
\normalsize

\usepackage{verbatim}
\usepackage{amsmath}
\usepackage{amssymb}
\usepackage{amsfonts}
\usepackage{latexsym}
\usepackage{amscd}
\usepackage{amsthm}
\usepackage{graphicx}
\usepackage{xcolor}
\usepackage{url}
\usepackage{soul}
\usepackage{cite}
\usepackage{algorithmic}
\usepackage{hyperref}
\newcommand{\Ac}{\mathcal{A}}
\newcommand{\calC}{\mathcal{C}}
\newcommand{\Nc}{\mathcal{N}}
\newcommand{\NN}{\mathbb{N}}
\newcommand{\Lc}{\widehat{\mathcal{L}}}
\newcommand{\Ld}{\widehat{\mathcal{M}}}
\newcommand{\Ic}{\mathcal{I}}
\newcommand{\Pc}{\mathcal{P}}
\newcommand{\FF}{\mathbb{F}}

\newcommand{\Mc}{\mathcal{M}}
\newcommand{\Yc}{\mathcal{Y}}

\newcommand{\rme}{\mathrm{e}}
\newcommand{\rmQ}{\mathrm{Q}}

\newtheorem{pro}{Proposition}

\DeclareMathAlphabet{\mathcal}{OMS}{cmsy}{m}{n}
\usepackage{soul}

\begin{document}

\title{\LARGE  Polar Coding for the Binary Erasure Channel  with Deletions}
\author{Eldho~K.~Thomas,
        Vincent Y. F. Tan, {\em Senior Member, IEEE}, 
				Alexander~Vardy, {\em Fellow, IEEE},
        and \\ Mehul~Motani, {\em Senior Member, IEEE} %
\thanks{E.~K.~Thomas is with the Institute of Computer Science, University of Tartu, Estonia, 51014 (email: eldhokt@gmail.com). V.~Y.~F.~Tan  and M.~Motani are with the Department of Electrical \& Computer Engineering, National University of Singapore, Singapore 117583 (emails: vtan@nus.edu.sg, motani@nus.edu.sg). A.~Vardy is with  Department of Electrical \& Computer Engineering,
University of California San Diego, La Jolla, CA 92093, USA and the  School of Physical \& Mathematical Sciences,
Nanyang Technological University, Singapore 637371 (email: avardy@ucsd.edu). } \thanks{This work is partially funded by a  Singapore Ministry of Education (MoE) Tier 2 grant (R-263-000-B61-112).}}

\markboth{IEEE COMMUNICATIONS LETTERS}%
{Draft paper}

\maketitle

\begin{abstract}
We study the application of polar codes in deletion channels by
analyzing the cascade of a binary erasure channel (BEC) and a deletion
channel. We show how polar codes can be used effectively on a BEC
with a single deletion, and propose a list decoding algorithm with a  cyclic redundancy check for
this case. The  decoding complexity is $O(N^2\log N)$, where $N$ is the blocklength of the
code. An important contribution is an  optimization of  the amount of redundancy added to minimize the overall error probability.  Our theoretical results are corroborated by numerical simulations which show that the list size can be reduced to one and the original message can be recovered with high probability as the length of the code grows.
\end{abstract}


\IEEEpeerreviewmaketitle


\vspace{-.05in}
\section{Introduction}
Polar codes, invented by Ar{\i}kan  \cite{arikan}, are the first provably capacity-achieving codes with low encoding and decoding complexity. Ar{\i}kan's presentation of polar codes includes a successive cancellation decoding algorithm, which generally does not perform as well as the state-of-the-art error-correcting codes at finite block lengths~\cite{Hassani}. 
To improve the performance of polar codes,  Tal and Vardy \cite{vardy} devised a list decoding algorithm. The initial work of Ar{\i}kan considers binary symmetric   memoryless channels. There have been attempts to study polar codes for other channels, e.g.,  the AWGN channel~\cite{abbe}. However, there are not many constructions of polar codes for channels with memory. See \cite{wang} and references therein.

The deletion channel is a canonical example of a  non-stationary, non-ergodic channel with memory. It deletes symbols arbitrarily and the positions of the deletions are unknown to the receiver. A survey by Mitzenmacher~\cite{mitzen} discusses the major developments in the understanding of deletion channels in greater detail.  To date, the Shannon capacity of deletion channels, in general, remains unknown. However, there have been attempts to find upper and lower bounds on the capacity of deletion channels \cite{ramji,Mit}.

Our motivation is partly  the work of Dolecek and Anantharam~\cite{dole}, in which the run length properties of Reed-Muller (RM) codes were exploited to correct a certain number of substitutions together with a {\em single} deletion; our work involves correcting {\em erasures} rather than substitiutions. RM codes and polar codes have similar algebraic structures and therefore polar codes are also potential candidates for correcting single deletions. However, they cannot be used directly on deletion channels since the polarization of a channel with memory has not been well-studied. Developing  polarization techniques for deletion channels is beyond the scope of this study. Instead, motivated by decoders that are possibly defective and delete symbols arbitrarily, we consider polar codes over a binary erasure channel (BEC) and an adversarial version of the deletion channel with one deletion, and provide a list decoding algorithm to successfully recover the original message with high probability\footnote{In this letter, we use the term w.h.p.\ to mean with probability tending to~$1$ as the blocklength  of the code $N$ tends to infinity.} (w.h.p.).  Unlike RM codes, polar codes do not have rich run length properties. Instead, we use the successive cancellation algorithm~\cite{arikan} for decoding. In addition, we provide a detailed analysis of the error probability, which was lacking in \cite{dole}. Channel cascades were studied previously in \cite{aaron} but our model has not been previously considered in the literature. We argue that the capacity of the cascade can be achieved; in constrast, \cite{dole} does not discuss capacity issues. 
 



\vspace{-.05in}
\section{Preliminaries}\vspace{-.05in}
\label{sec:prelim}

\vspace{-.05in}
\subsection{Polar Codes}\vspace{-.05in}
\label{sec:twoa}
We consider polar codes of length $N=2^n$ constructed recursively 
from the kernel 
$G_2= \genfrac{(}{)}{0pt}{}{1\,0}{1\,1}$.
Given an information  vector (message) $u_1^N=(u_1,\ldots,u_N)$ where $u_i \in \FF_2$, a codeword $x_1^N$ is generated using the relation $x_1^N=u_1^NB_NG_2^{\otimes n}$ where $G_2^{\otimes n}$ is the $n$-th Kronecker product  of $G_2$ and $B_N$ is a bit-reversal permutation matrix, defined explicitly in \cite{arikan}.
The vector
$x_1^N$ is transmitted through $N$ independent copies of a binary discrete memoryless channel (BDMC) $W: \FF_2 \rightarrow \Yc$  with transition probabilities $\{W(y|x):x\in\FF_2,y\in {\cal Y}\}$ and capacity $C(W)$. As $n$ grows, the individual channels start polarizing. That is,  a subset of the channels tend to noise-free channels and others tend to completely noisy channels. The fraction of noise-free channels tends to the capacity $C(W)$. The polarization behavior suggests using the noise-free channels to transmit information bits, while setting the inputs to the noisy channels to values that are known \emph{a priori} to the decoder (i.e., the frozen bits). That is, a message  vector $u_1^N$ consists of information bits and frozen bits (often set to zero) where  $\Ic \subset \{1,\ldots, N\}=\Nc$ of size $k$ is the information set and $\bar{\Ic}$ is the set of frozen bits. This scheme achieves capacity~\cite{arikan}. Denote the channel output by $y_1^N=(y_1,\ldots,y_N)$ and the $i$-th synthesized subchannel  with input $u_i$ and output $(y_1^N, u_1^{i-1})$ by $W_N^{(i)}$ for $i=1,\ldots ,N$. The transition probability  matrix $W_N^{(i)}$ is defined as 
\begin{equation}
W_N^{(i)}(y_1^N  , u_1^{i-1}|u_i):= \sum_{u_{i+1}^N \in\mathbb{F}_2^{N-i}  }\frac{1}{2^{N-1}}W_N(y_1^N|u_1^N),\nonumber
\end{equation}
where $W_N(y_1^N| u_1^N)\!:=\!\prod_{i=1}^N W(y_i|x_i)$ and $x_1^N=u_1^NB_NG_2^{\otimes n}$ is the codeword corresponding to the message $u_1^N$. The encoding complexity of polar coding is  $O(N \log N)$~\cite{arikan}. 

\subsection{Successive Cancellation Decoding}
\label{sec:twob}
Ar{\i}kan \cite{arikan} proposed a successive cancellation (SC) decoding scheme for polar codes. Given $y_1^N$ and the estimates $\hat{u}_1^{i-1}$ of $u_1^{i-1}$, the SC algorithm estimates $u_i$. The following logarithmic likelihood ratios (LLR) are used to estimate each $u_i$ for $ i =1,\ldots , N$:
\begin{eqnarray*}
L_N^{(i)}(y_1^N, {\hat{u}_1^{i-1}})=\log  \frac{W_N^{(i)}(y_1^N, \hat{u}_1^{i-1}|u_i=0)}{W_N^{(i)}(y_1^N, \hat{u}_1^{i-1}|u_i=1)}.
\end{eqnarray*}
The estimate of an unfrozen bit $u_i$ is determined by the signs of the LLRs, i.e.,  $\hat{u}_i  = 0$ if $L_N^{(i)}(y_1^N, {\hat{u}_1^{i-1}})\ge 0$ and $\hat{u}_i=1$ otherwise.  
It is known that polar codes with SC decoding achieve capacity with decoding complexity of $O(N \log N)$ \cite{arikan}. 

\subsection{Adversarial Deletion Channel}\label{sec:adv}
 We suppose that $N$  bits are sent over a channel and exactly $d$ bits are deleted. We call this a {\it $d$-deletion channel}. That is, for $N$ bits sent, the decoder only receives $N-d$ bits after $d$ deletions and the positions of deletions are not known to the receiver.   Note that  this is not the   probabilistic deletion channel in which each symbol is independently deleted with some fixed probability $q\in (0,1)$~\cite{Mit}.


\section{Problem Setting and Model}
\label{sec:pblm}
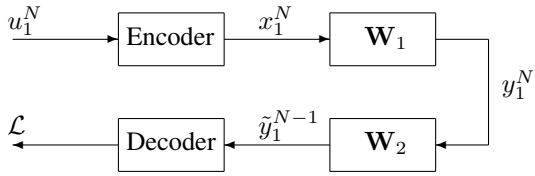
\begin{figure}
\centering
\begin{picture}(210, 60)
\put(3,54){\mbox{$u_1^N$}}
\put(98,54){\mbox{$x_1^N$}}
\put(98,14){\mbox{$\tilde{y}_1^{N-1}$}}
\put(3,14){\mbox{$\mathcal{L}$}}
\put(190,30){\mbox{$ {y}_1^{N }$}}

\put(48,48){\mbox{Encoder}}
\put(48,8){\mbox{Decoder}}

\put(138,48){\mbox{$\mathbf{W}_1$}}
\put(138,8){\mbox{$\mathbf{W}_2$}}

\put(5,50){\vector(1,0){40}}
\put(45,60){\line(1,0){40}}
\put(45,40){\line(1,0){40}}
\put(45,40){\line(0,1){20}}
\put(85,40){\line(0,1){20}}

\put(85,50){\vector(1,0){40}}
\put(125,60){\line(1,0){40}}
\put(125,40){\line(1,0){40}}
\put(125,40){\line(0,1){20}}
\put(165,40){\line(0,1){20}}

\put(165,50){\line(1,0){20}}
\put(185,50){\line(0,-1){40}}
\put(185,10){\vector(-1,0){20}}
\put(125,20){\line(1,0){40}}
\put(125,00){\line(1,0){40}}
\put(125,00){\line(0,1){20}}
\put(165,00){\line(0,1){20}}

\put(125,10){\vector(-1,0){40}}
\put(45,20){\line(1,0){40}}
\put(45,00){\line(1,0){40}}
\put(45,00){\line(0,1){20}}
\put(85,00){\line(0,1){20}}
\put(45,10){\vector(-1,0){40}}
\end{picture}
\caption{BEC-1-Deletion Cascade.  $\mathbf{W}_1=\mathrm{BEC}(p)^N$ is the length-$N$ BEC, $\mathbf{W}_2$ is the $1$-deletion channel, and $\mathcal{L}$ is the list of possible messages.}
\label{model}
\end{figure}

Consider the {\it 1-deletion channel}  ($d=1$ in the  definition in Section~\ref{sec:adv}), where exactly one bit is  deleted.  We suppose  that $N=2^n$ where $ n \in \NN$. A  message vector $u_1^N$ is encoded using the polar encoder   and is sent across $N$ uses of a   BEC $W_1^N= {\bf W}_1$, each with erasure probability $p\in (0,1)$. The output vector is passed through  a 1-deletion channel ${\bf W}_2$. We denote this cascade of $\mathbf{W}_1$ and $\mathbf{W}_2$ as $\mathbf{W}$ and call this a \emph{BEC-1-Deletion Cascade}.  This model is shown in Fig.~\ref{model}.  The output of $\mathbf{W}$ is denoted as $\tilde{y}_1^{N-1}$. Note that  $\mathbf{W}$ permits erasures and a single deletion.  That is, a message $u_1^N$ is sent across $\mathbf{W}$ and a vector $\tilde{y}_1^{N-1}$ is received. A decoder is designed in such a way that w.h.p., a list $\mathcal{L}$ (of linear size in $N$) containing an estimate $\hat{u}_1^N$ of the original message $u_1^N$ is returned. 



\section{Coding for the BEC-1-Deletion Cascade}
\label{sec:cas}

\subsection{Reconstruction of the BEC Output}
\label{sec:decoding}
A message $u_1^N$ is sent over a BEC-1-Deletion cascade using a polar encoder described in Section~\ref{sec:twoa} and $\tilde{y}_1^{N-1}$ is received. In order to decode $\tilde{y}_1^{N-1}$, we use the SC algorithm (refer to Section~\ref{sec:twob}). Since the position of the deletion is unknown, we first identify a set of vectors, called the {\em candidate set}, which contains $\tilde{y}_1^{N-1}$ as a sub-sequence.  A na\"ive algorithm to construct the candidate set would be to insert $0, 1, \rme$ in the $N$  locations before and after each symbol of $\tilde{y}_1^{N-1}$. We then apply the  SC algorithm to each vector in the candidate set. 

For example, suppose $N=4$ and the received vector is $\tilde{y}_1^3= 01\rme$. Then the following set $\mathcal{S}$ includes all vectors which contain the subsequence $01\rme$:
\begin{align*}
\mathcal{S}=\left\{001\rme, 101\rme, \rme01\rme, 011\rme, 0\rme 1\rme, 010\rme,01\rme\rme,  01\rme0, 01\rme1  \right \}
\end{align*}
The size of this set can be further reduced if we notice that inserting $\rme$ at $N$ positions is enough to identify all possible messages those can output $\tilde{y}_1^{N-1}$ after a single deletion.
This is because of the following: Suppose the $i$-th symbol is deleted from $y_1^N$. Instead of inserting 0 or 1 at position $i$, we insert an erasure symbol $\rme$. Since a polar code correcting $\alpha \approx Np' $ (where $p'<p$)  erasures also corrects $\alpha+1$ erasures w.h.p., under the SC decoding algorithm, this new length-$N$ vector decodes to the correct message w.h.p.\ no matter which symbol was at position $i$. We state this observation formally:

\begin{pro}
\label{pro:cand}
Suppose $u_1^N$ is sent over a BEC-1-Deletion cascade ${\bf W}$. (See Fig.~\ref{model}.)  
The size of the candidate set $\mathcal{A}$ (constructed above) is $N-\alpha$ where $\alpha$ is the number of erasures present in the received string $\tilde{y}_1^{N-1}$.
\end{pro}
\begin{IEEEproof}
The candidate set is $$\Ac= \{(\tilde{y}_1^{i-1},\rme, \tilde{y}_{i}^{N-1}): ~ i= 1,2,\ldots, N  \} \subset \{0,1,\rme\}^N$$
where $\tilde{y}_1^{N-1}$ is the received string. 
Suppose that the  $j$-th symbol of $\tilde{y}_1^{N-1}$ is $\rme$. Inserting another $\rme$ before the  $j$-th symbol $\rme$ forms vector $\tilde{y}_1^{j-1}\rme\rme\tilde{y}_{j+1}^{N-1}$. This vector repeats if we insert $\rme$ again after the the $j$-th symbol $\rme$.
Therefore, considering  non-erasure bits of  $\tilde{y}_1^{N-1}$ and inserting  exactly one  erasure symbol $\rme$ at  positions before and after these non-erasure bits produces unique vectors in the candidate set $\mathcal{A}$.  
Since the number of erasure symbols is $\alpha$, the total number of vectors in $\mathcal{A}$ is $N-\alpha$.
\end{IEEEproof}

We remark that as $N \to \infty$,  by the law of large numbers $\frac{\alpha}{N} \to p$ and hence $|\mathcal{A}| \approx N-Np$ where $p \in (0,1)$ is the erasure probability of the BEC.  

\subsection{List Decoding}
After the construction of the set $\mathcal{A}$, the problem  reduces to the  decoding of each vector in $\mathcal{A}$ using the SC algorithm. Since $|\mathcal{A}|= N-\alpha$, we get a list of messages of size at most $N-\alpha$ at the end of the whole decoding procedure.

Let $\mathrm{SC}(y_1^N)$ denote the SC decoding of $y_1^N$, and define
\begin{equation}
\label{eq:listl}
\mathcal{L}= \{{u}_1^k: u_1^k=\hat{u}_1^N|_{\Ic}, \hat{u}_1^N = \mathrm{SC}(y_1^N), y_1^N \in \mathcal{A}\},
\end{equation} 
as the list of messages returned by the set $\mathcal{A}$ where $\Ic $ is the information set.  

Since we insert the erasure symbol $\rme$ at each of the $N$ possible positions  (including the deleted position), the original message sent  belongs to  $\mathcal{L}$ w.h.p. 
Ar{\i}kan~\cite{arikan} proved that the probability of error $P_{\rme}^{(N)}$ vanishes asymptotically for polar codes over any BDMC. A more precise estimate was provided by Ar{\i}kan and Telatar~\cite{telatar} who showed that for any $\beta \in (0,1/2)$,  $P_{\rme}^{(N)} \le 2^{-N^\beta}$   for sufficiently large block lengths $N$.  Therefore, under   SC decoding, vectors in  $\mathcal{A}$ return all possible messages that can produce the string  $\tilde{y}_1^{N-1}$ under a   single (adversarial) deletion.

\subsection{Recovering the Correct Message from the List  via Cyclic Redundancy Check (CRC)}\label{sec:crc}
Naturally, there can be multiple $u_1^k \in \Mc$ that belong to the list $\mathcal{L}$ and it may  not be easy to single out the original message. However, by applying a simple pre-coding technique  using an $r$-bit CRC (or a code having an $r\times k$ parity check matrix) \cite{chen, vardy}, the original message can be detected from the list,  albeit with some additional probability of error.  We describe how to recover the correct message w.h.p.\ here.


Recall that we have $N-k$ frozen bits that we usually set to zero. Instead of setting all of them to zero, we set $N-k-r$ frozen bits to zero, where $r$ is a small number we optimize in Section~\ref{sec:analysis}. These $r$ bits will contain the $r$-bit CRC value of the $k$ unfrozen bits (or simply the parity bits). To generate a $r$-bit CRC, we select a polynomial of degree $r$, called a {\em CRC polynomial}, having  $r+1$ coefficients. We then divide the message (by treating it as a binary polynomial) by this CRC polynomial to generate a remainder of degree at most $r-1$, with total number of coefficients $r$. We append these $r$ coefficients  at the end of the $k$-bit message to generate a $(k+r)$-bit vector. To verify that the correct message is received, we perform the polynomial division again to check if the remainder is zero. For more details on the choice of CRC polynomials, please refer to \cite{koopman}.  We send these $k+r$ bits  across the cascade. This new encoding is a slight variation the original polar coding scheme \cite{arikan}. Also, note that the original information rate $R= \frac{k}{N}$ is preserved. However, the  rate of the polar code is slightly increased to $R_{\mathrm{polar}}= \frac{k+r}{N}$. 

To summarize, we encode the message $u_1^{k}$ of length $k$ into a length $k+r$ vector $u_1^{k+r} \in \calC'$ having redundancy $r$ where $|\calC'| = 2^{k}$. Then we apply the polar coding scheme for the codebook $\calC'$. This will result in a polar code $\calC$ of length $N$ and  size $2^{k+r}$ where only the subset $\calC' \subset \calC$ carries information that we wish to transmit. The codeword $x_1^N \in \calC$ corresponding to the original message $u_1^{k} $ is then passed through the BEC-1-Deletion  channel and outputs a vector $\hat {y}_1^{N-1}$. After constructing the set $\mathcal{A}$ by inserting $e$ at each possible $N$ positions, we apply the SC algorithm on $\mathcal{A}$. However, not all of these resulting vectors in $\calC$ carry information. We can check this using the initial $r$-bit CRC (or the parity check matrix). All vectors which fail under the CRC check are removed and we then select the message with the maximum likelihood from the list.

\subsection{Analysis and Optimization of the Overall Error Probability}\label{sec:analysis}
Suppose $H$ denotes the $r\times (k+r)$ parity check matrix with rows $\{h_i :i=1,\ldots, r\}$ that is being used for adding parity to the $k$ bit message. Then the set of messages that carries any information  can be identified as
$$\Ld:= \big\{u_1^{k}: u_1^{k+r}H^T=0, u_1^{k+r} \in \Lc\big\},$$ where $\Lc$ is the modified version of (\ref{eq:listl}) according to the new polar coding scheme defined as 
$$\Lc:= \{{u}_1^{k+r}: u_1^{k+r}=\hat{u}_1^N|_{\Ic \cup \Pc}, \hat{u}_1^N = \mathrm{SC}(y_1^N), y_1^N \in \mathcal{A}\},$$ and where $\Pc \subset \bar{\Ic}$ is the set  of parity bits ($\bar{\Ic}$ is the set of frozen bits). 
If the rows of $H$ are chosen uniformly and independently  from 
$\{0,1\}^{k+r}$, 
the probability that a vector $u_1^{k}$ is in  $\Ld$ is $$\Pr\big(u_1^{k} \in \Ld \,\big)= \Pr \left(\langle h_i, u_1^{k+r}\rangle = 0, \,\forall\,  i=1,\ldots, r \right)= \frac{1}{2^r}$$ where $ u_1^{k+r} \in \Lc$. That is, a message in $\Lc$ is wrongly identified as the original message with probability $1/2^r$. However, the true message sent satisfies the parity-check condition $u_1^{k+r}H^T=0$.  Therefore, by the union bound, the total probability that an incorrect message is returned is  upper bounded as
\begin{equation}
\label{eq:err}
P_{\mathrm{TotErr}}^{(N)}\le \frac{|\Lc|}{2^r}+ |\mathcal{A}|P_{\rme}^{(N)},
\end{equation}
 where $P_{\rme}^{(N)}$ is the probability of error of the SC decoding algorithm and $|\Lc| \le |\mathcal{A}| \approx N(1-p)$ for a single deletion.

To maintain that $R_{\mathrm{polar}} \approx R$ (that is, as the block length $N$ grows, $R_{\mathrm{polar}}$ converges to $R$) and the upper bound on $P_{\mathrm{TotErr}}^{(N)}$ in (\ref{eq:err}) is minimized, we have to choose $r$ carefully.

For a single deletion, the size of the candidate set $|\mathcal{A}|\approx N(1-p)$ and hence $|\Lc|\le N(1-p)$ w.h.p. From Hassani {\em et al.} \cite{hassa}, the rate-dependent error probability 
 of the polar code for the BEC with rate $R_{\mathrm{polar}}$ is 
$$P_{\rme}^{(N)} = 2^{-2^{\frac{n}{2}+ \frac{\sqrt{n}}{2}\rmQ^{-1}\left(\frac{R_{\mathrm{polar}}}{C({\bf W})}\right)+o(\sqrt{n})}}.$$
where $N=2^n$, $\rmQ(x):=\frac{1}{\sqrt{2 \pi}}\int_{x}^{\infty} \exp(-\frac{t^2}{2})\, \mathrm{d}t$ is the complementary Gaussian cumulative distribution function, and $C({\bf W})$ is the capacity of the channel cascade.

From (\ref{eq:err}), 
\begin{align}
P_{\mathrm{TotErr}}^{(N)} \le   |\mathcal{A}|\bigg[2^{-r} + 2^{-2^{\frac{n}{2}+ \frac{\sqrt{n}}{2}\rmQ^{-1}\left(\frac{R_{\mathrm{polar}}}{C({\bf W})}\right)+o(\sqrt{n})}}\bigg ]\label{eqn:two_terms} .
\end{align} 
It can be verified easily that the first term  in the square parentheses in \eqref{eqn:two_terms} is decreasing and the second term with $ R_{\mathrm{polar}}=\frac{k+r}{N}$ is increasing in $r$. To optimize the upper bound in \eqref{eqn:two_terms}, we set the exponents of two terms to be   equal (neglecting the insignificant $o(\sqrt{n})$ term), i.e.,
$$r= 2^{\frac{n}{2}+ \frac{\sqrt{n}}{2}\rmQ^{-1}\left( \frac{k+r}{NC({\bf W})}\right)}= \sqrt{N}2^{\frac{\sqrt{\log_2 N}}{2}\rmQ^{-1}\left( \frac{k+r}{NC({\bf W})}\right)},$$
where we used the fact that $N=2^n$.

Now we find an expression for $r$ in terms of the backoff from capacity.
To transmit the code at a rate close to the capacity, for a small constant $\delta>0$, assume that $R=  (1-\delta)(1-p)$ where $C({\bf W})= 1-p$ since a polar code over the BEC 1-deletion cascade achieves the capacity of the BEC; this is a simple consequence of~\cite[Problem~3.14]{networkIT} and the fact that the list size is polynomial. Then the rate $R_{\mathrm{polar}}= R+ \frac{r}{ N}(1-p) \ge (1-\frac{\delta}{2})(1-p)$ for $N$ large enough. Therefore, 
$$r= \sqrt{N}\cdot 2^{\frac{\sqrt{\log_2 N}}{2}\rmQ^{-1}\left( 1-\frac{\delta}{2} \right)}.$$
Let $z= \rmQ^{-1}(1-\frac{\delta}{2})$. Since $\frac{\delta}{2} \approx 0$, $z\ll 0$.
Then $\rmQ(z) = 1-\frac{\delta}{2}$ and hence $\frac{\delta}{2}=\rmQ(-z)$.
Since $\rmQ(-z)$ decays as $e^{-z^2/2}$  as $z\to-\infty$, $z^2 = 2 \ln{\frac{2}{\delta}}$.  
Then $z= -\sqrt{2 \ln{\frac{2}{\delta}}}$. Therefore, the optimal value of the number of parity bits $r$ is
$$r= \sqrt{N}\cdot 2^{-\sqrt\frac{(\log_2 N) (\ln{\frac{2}{\delta}})}{2}} = \Theta(\sqrt{N}).$$
This is a rate-dependent choice of $r$ (through $\delta$) that simultaneously ensures that $R_{\mathrm{polar}}\to R$ and the upper bound on $P_{\mathrm{TotErr}}^{(N)}$ in (\ref{eq:err}) is minimized.

 \vspace{-.1in}

\subsection{Finite Number of Deletions}
Now consider the cascade of a BEC and a $d$-deletion channel where $d\in\mathbb{N}$ is finite. This model can   be analyzed using the same techniques presented here. The only difference is the size of the candidate set $\mathcal{A}$. By using the same arguments as  in the $1$-deletion case, we construct $\mathcal{A}$  by inserting erasure symbols at $d$ positions and  $|\mathcal{A}|={N \choose d}-\alpha$. 
Therefore, the list size $|\Lc| \le {N \choose d}-\alpha$. Since the models are similar, a CRC construction and error probability analysis for the BEC-$d$-Deletion cascade similar to that presented in Sections~\ref{sec:crc} and~\ref{sec:analysis} respectively can be performed. In addition, we see that even if the list size is  $d= o\big(\frac{N}{\log  N}\big)$, the capacity of the BEC  is achieved because   $|\Lc| \le N^d$ is still subexponential.

\subsection{Complexity of the Decoding Algorithm}
The encoding  complexity of the BEC-1-Deletion cascade is same as that for standard polar codes, i.e., $O(N\log N)$. However, the SC decoding algorithm has to be applied  to   all vectors in the candidate set $\mathcal{A}$ of size $N-\alpha$ (cf.\ Prop.~\ref{pro:cand}).  Thus, 
the complexity of the decoding algorithm of the BEC-1-Deletion cascade is  $O(N^2 \log N)$ and that for the BEC-$d$-Deletion cascade is $O(N^{d+1} \log N)$.
Although the complexity of the decoding algorithm increases    by $O(N)$   for each additional deletion, it can still be performed in polynomial time. 

\section{Simulation Results}\label{sec:sim}
In this section, we demonstrate the utility of the proposed algorithm by performing numerical simulations. 
 The simulations are carried out in \textsc{Matlab}  using code provided  in~\cite{vangala} with the following parameters.\footnote{The \textsc{Matlab} code to reproduce the simulations is provided at \url{https://www.ece.nus.edu.sg/stfpage/vtan/commL_code.zip}.}  Let $n=\log_2 N$ vary from $6$ to~$11$. The erasure probability of the BEC is $p=0.3$. Thus, the capacity of the cascade is $C(\mathbf{W})=0.7$. We consider three different code rates: $R=0.50, 0.55$ and $0.60$. We fix $r= \lceil 0.7\sqrt{N} \rceil$    and the $r$-bit CRC polynomial is   chosen according to~\cite{koopman}. The error probability is computed by averaging over $1000$ independent runs.  
 
\begin{figure}[t]
\centering
\includegraphics[width = .9\columnwidth]{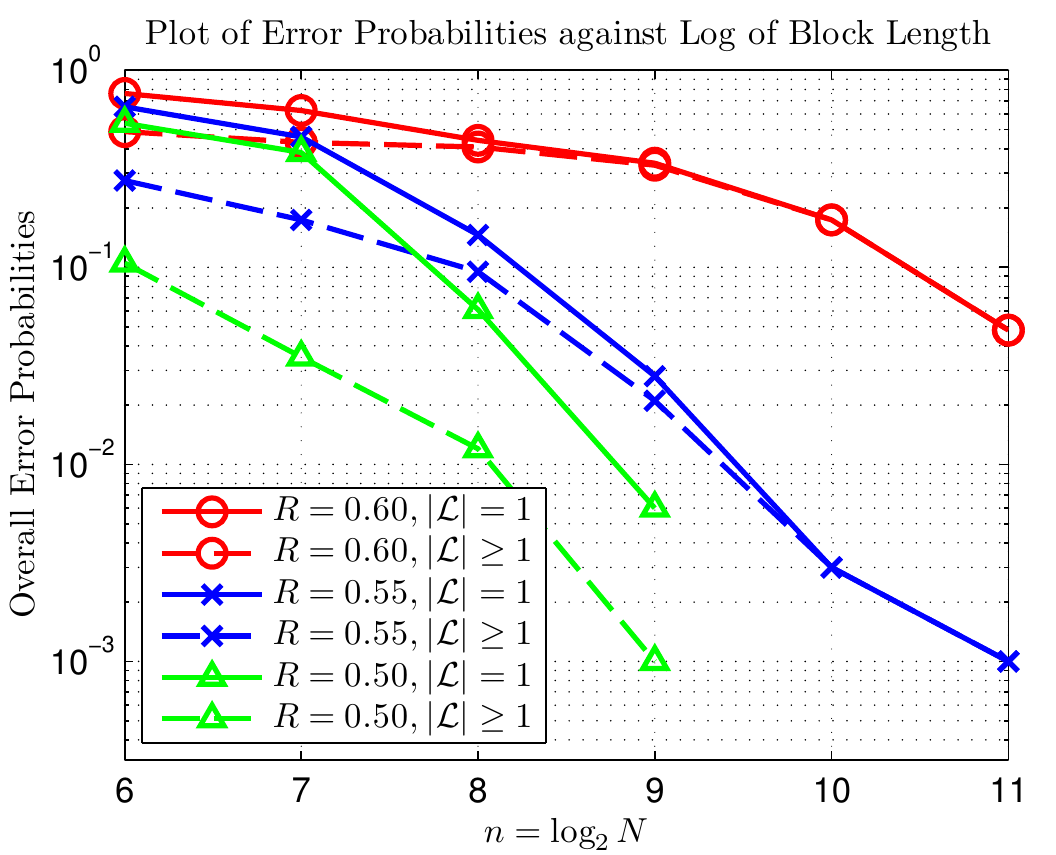}
\caption{Plot of error probabilities against $n=\log_2 N \in \{6,\ldots, 11\}$.  The solid line is the error probability  of obtaining a list of size $1$, which is exactly the original message. The broken line is the error  probability  of obtaining a list of size at least $1$ containing the original message. The list size after using    CRC  is small even though it is  not exactly $1$. Data points that are not available indicate that the simulated error probability over $1000$ runs is  exactly $0$.}
\label{fig:errprob}
\end{figure}

 We encode a   random length-$\lceil RN\rceil$  message using a $r$-bit CRC polynomial so that the input of the encoder is a $k +r$ length input vector and the output is an $N$-bit vector. This vector  is then transmitted through a BEC-1-deletion cascade and received a length-$(N-1)$ vector. The CRC list decoder then computes a list  of possible messages given the channel output.   Fig.~\ref{fig:errprob}  shows  that, with a suitable choice of the number of CRC bits $r$ and CRC polynomials, as $N$ grows, the list is of size $1$ and contains only the original message w.h.p.  

\bibliographystyle{unsrt}

\end{document}